\documentstyle[prl,aps,epsf,multicol]{revtex} 
\begin{document}
%\twocolumn
\title{Upper Bounds on the Superfluid Stiffness of Disordered Systems}
\author{Arun Paramekanti\thanks{arun@theory.tifr.res.in}, Nandini Trivedi
\thanks{ntrivedi@theory.tifr.res.in}, and Mohit Randeria
\thanks{randeria@theory.tifr.res.in}}
\address{Theoretical Physics Group, Tata Institute of Fundamental 
Research, Mumbai 400005, India}
\maketitle 
%\begin{minipage}[t]{6.0in}
%\address{ 
\begin{abstract}
We derive several upper bounds for the superfluid 
stiffness $D_s$ for Bose and Fermi systems in terms of expectation 
values of local operators using linear response theory
and variational methods. 
These give insight into the non-trivial dependence of $D_s$
on parameters such as disorder and interaction in
systems with broken continuous translational invariance. 
Our best variational bound for disordered systems is obtained by 
allowing the phase twist applied at the
boundary to be distributed inhomogeneously within the system.
Path integral quantum Monte Carlo simulations are used to
quantitatively compare the bounds and $D_s$ 
for disordered interacting Bose systems.
%\typeout{polish abstract}
\end{abstract}
%\pacs{PACS numbers: }
%\end{minipage}
%\pacs{PACS numbers: }
%\newpage
%\narrowtext
%%%%%%%%%%%%%%%%%%%%%%%%%%%%%%%%%%%%%%%%%%%%%%%%%%%%%%%%%%%%%%%%%%

\newcommand{\be}{\begin{equation}}
\newcommand{\den}{\overline{n}}
\newcommand{\ee}{\end{equation}}
\newcommand{\bea}{\begin{eqnarray}}
\newcommand{\eea}{\end{eqnarray}}
\newcommand{\nn}{\nonumber}
\newcommand{\la}{\langle}
\newcommand{\ra}{\rangle}
\newcommand{\th}{\theta}
\newcommand{\ph}{\phi}
\newcommand{\dg}{\dagger}
\renewcommand{\vr}{{\bf{r}}}
\newcommand{\hj}{\hat{\alpha}}
\newcommand{\hx}{\hat{1}}
\newcommand{\crx}{c^\dg(\vr)c(\vr+\hx)}
\newcommand{\crkubox}{c^\dg(\vr)c(\vr+\hat{x})}
\newcommand{\crj}{c^\dg(\vr)c(\vr+\hj)}
\newcommand{\crmj}{c^\dg(\vr)c(\vr - \hj)}
\newcommand{\sumall}{\sum_{\vr}}
\newcommand{\sumx}{\sum_{r_1}}
\newcommand{\nabj}{\nabla_\alpha \theta(\vr)}
\newcommand{\nabx}{\nabla_1 \theta(\vr)}
\newcommand{\sumy}{\sum_{r_2,\ldots,r_d}}
\newcommand{\krj}{K(\vr,\vr+\hj)}
\newcommand{\sigr}{|\psi_0\rangle}
\newcommand{\sigl}{\langle\psi_0 |}
\newcommand{\sier}{|\psi_{\Phi}\rangle}
\newcommand{\siel}{\langle\psi_{\Phi}|}
\newcommand{\sumrj}{\sum_{\vr,\alpha=1\ldots d}}
\newcommand{\krw}{K(\vr,\vr+\hx)}
\newcommand{\Dtheta}{\Delta\theta}
\newcommand{\rhonew}{\hat{\rho}(\Phi)}
\newcommand{\rhoold}{\hat{\rho_0}(\Phi)}
\newcommand{\dt}{\delta\tau}
\newcommand{\cP}{{\cal P}}
\newcommand{\cS}{{\cal S}}
\newcommand{\hnr}{\hat{n}({\vr})}
%\begin{multicols}{1}

\bigskip

\noindent
{\bf I. Introduction:} 
\medskip
%\twocolumn

The superfluid density $n_s$, or the closely
related stiffness $D_s$ (defined below),
characterizes the phase rigidity of a superconductor or a superfluid.
Experimentally $n_s$  is more easily measured
than the order parameter since external probes, such as magnetic fields
or rotation,
can couple to the relevant correlation function. Further, even
from a theoretical point of view, $n_s$ is arguably the more fundamental 
characterization: e.g., in two dimensional systems $n_s$ can be non-zero 
even though the order parameter vanishes at finite temperatures.

In translationally invariant (one component) superfluid systems, 
it can be shown on general grounds that at $T=0$, $n_s= n$, the total
density: all of the fluid participates in superflow \cite{pines}.
This follows from translational invariance and the absence of low lying
excitations that can couple to a transverse probe. 
However, as soon as one destroys continuous
translational invariance, either by the presence of a lattice
or impurities, $n_s$ is no longer constrained to be the total
density $n$, and becomes smaller as we shall describe below. 
The dependence of $n_s$ at $T=0$ on
disorder is of considerable interest since with increasing disorder
one could suppress the superfluid density and eventually drive the
system into a non-superfluid, possibly insulating
state \cite{fisher,sc}. 

In this paper we will derive exact upper bounds on $n_s$
(or $D_s$) which help understand how the superfluid density 
varies with interaction and disorder in both Fermi and Bose systems. 
An outline of the paper and a summary of our main results is as follows.
Section II introduces the notation and definitions used in the rest
of the paper. In Section III we discuss bounds for $D_s$ obtained from
linear response theory, emphasizing the effect of breaking
continuous translational invariance. Two simple upper bounds, 
eqn.~(\ref{ke-bd}) and eqn.~(\ref{den-bd}), are derived and compared
with results of path integral quantum Monte Carlo (QMC) simulations for
non-disordered lattice boson problems.

In Section IV we derive variational upper bounds on $D_s$ for a system
with disorder by constructing a trial wave 
function for a system with twisted boundary conditions given the exact 
ground state for the untwisted case.
We obtain two bounds: one, $D^o_s$ given in eqn.~(\ref{leg-bd}),
which is the lattice analog of a result derived earlier by Leggett
\cite{legg1} for continuum systems, and
an improved variational bound $D^{\ast}_s$ given in eqn.~(\ref{var-bd}).
The calculation of $D^{\ast}_s$
is shown to be equivalent to calculating the
conductance of a resistor network problem as explained in the discussion 
following eqn.~(\ref{var-bd}). We also briefly discuss the
generalization of these results to finite temperature.

The physical idea underlying the bounds $D^o_s$ and $D^{\ast}_s$
is the following. In a homogeneous system the applied twist at the 
boundaries is distributed uniformly across the entire system. 
However, in an inhomogeneous system the twist is accommodated preferentially
in those regions of space where it costs the least energy, leading to
a lower stiffness.
Our results can be summarized by the following set of inequalities
\be
D_s \leq D^{\ast}_s \leq D^o_s \leq \pi \la K_1 \ra \leq 2 t \pi \den
\label{results}
\ee
where $D_s$ is the exact superfluid stiffness, 
$D^o_s$ and $D^{\ast}_s$ are the bounds mentioned above, 
$\la K_1 \ra$ is (the absolute value of) 
the mean kinetic energy (KE) on a bond, 
$t$ is the hopping strength in the lattice
model and $\overline{n}$ is the filling or the average 
particle density.

In Section V we illustrate the quantitative usefulness of our results by
discussing various examples and comparing the bounds
with $D_s$ obtained from QMC calculations on 
disordered Bose systems. 
(The details of the QMC are given in the Appendix.)
The bounds are found to capture the
non-trivial parameter dependence of $D_s$, including its nonmonotonic
variation in some cases. However, these bounds, which are based on local
correlation functions, are not in general sensitive to the superfluid-insulator transition.
In Section VI, we make several remarks about our results.
Amongst other things we briefly discuss charge vs superfluid stiffness,
describe a model for which $D^{\ast}_s$ captures a percolation transition 
that is missed by the other bounds, and obtain bounds on $T_c$ for 2D 
systems. We summarize our results in Section VII.

\bigskip

\noindent
{\bf II Definitions and Notation}

\medskip

We consider interacting Bose or Fermi systems
with a generic Hamiltonian
\be
\hat{H}=-\sumrj t_\alpha(\vr) \left[ \crj + h.c. \right] + \hat{V}.
\label{hamiltonian}
\ee
We work on a hypercubic lattice in $d$ dimensions with lattice spacing 
$a=1$, $\vr$ runs over all $N=L^d$ lattice sites and $\hj$ denotes the 
unit vector along the $\alpha^{\rm th}$ direction. 
$c$ and $c^{\dg}$ are annihilation and creation operators; 
for Fermi systems we may also need spin labels which are omitted for
simplicity.
$\hat{V}$ includes all possible interactions between
the particles as well as the possibly random (one-body) potential and is an
operator which is diagonal in the number basis. Finally,
$t_\alpha(\vr)$ is the hopping amplitude between nearest neighbour 
sites $\vr$ and $\vr+\hj$.
This may again be a random function but is restricted to be positive. 
(Note that we do {\it not} consider frustration introduced by having both 
positive and negative hopping amplitudes.)

The kinetic energy operator on the link $(\vr, \vr+\hj)$ is
defined as
\be
K_\alpha(\vr)= t_\alpha(\vr) \left[\crj+h.c.\right].
\label{ke-op}
\ee
For later convenience, this definition differs from the standard one
by a negative sign. The paramagnetic current along $\hat{\alpha}$ at 
point $\vr$ is given by
\be
j_\alpha(\vr)= -i t_\alpha(\vr) \left[\crj - h.c.\right].
\ee
The (paramagnetic) current-current correlation function for
an $N = L^d$ site system at zero temperature and zero frequency is given by
\be
\chi_1({\bf q}, \omega=0)= \frac{2}{N}\sum_{n\neq 0}
\overline{\frac{|\la \psi_n| j_1({\bf q})|\psi_0\ra|^2}{\omega_{n}-
\omega_{0}}}
\label{jj}
\ee
where $|\psi_n\ra$ is the $n^{th}$ eigenstate of the system with energy
$\omega_n$ ($n=0$ denotes the ground state) and ${\bf q}$ is the momentum.
The overbar denotes an average over the quenched disorder.
(This is a shorthand notation for first evaluating the correlation 
function in real space, then disorder averaging and finally
Fourier transforming to momentum space.)
Now, within the Kubo formalism, we can define the superfluid stiffness $D_s$
in terms of the transverse current-current correlation \cite{baym,swz}:
\be
\frac{D_s}{\pi}= \la\psi_0|K_1|\psi_0\ra - 
\chi_1(q_1=0,{\bf q}_{\perp}\rightarrow 0, \omega=0)
\label{kubo}
\ee
where ${\bf q}_{\perp}=\{q_2,\ldots,q_d\}$ is the set of momentum
coordinates orthogonal to the $\hat{1}$
direction. The first term 
$\la K_1 \ra \equiv L^{-d}\sum_{\vr}\la\psi_0|K_1(\vr)|\psi_0\ra$ 
comes from the diamagnetic part of the response \cite{footnote}.

An equivalent definition of the superfluid stiffness is arrived at
by using twisted boundary conditions (b.c.) on the ground state 
wave function \cite{byers,jasnow}. This can be imposed in the
standard way by considering the system as a ``ring'' in the $\hx$
direction, threaded by a flux $\Phi$.
(Note that $\Phi=0$ corresponds to periodic b.c.'s).
In the site-occupation basis, moving one particle around the ring, 
keeping the others fixed, amounts to
changing the configuration from $\{ n \} \equiv
\{n(\vr_1),\ldots,n(\vr_m),\ldots,n(\vr_N)\}$      
with $n(\vr_m) \geq 1$, to
$\{ n^\prime \} \equiv
\{n(\vr_1),\ldots,n(\vr_m)-1,\ldots,n(\vr_N);
n(\vr_m+L\hx)=1\}$.
The twisted b.c. is then given by 
\be
\la \{n^\prime \} | \psi_\Phi \ra = e^{i\Phi}
\la \{n\} | \psi_\Phi \ra .
\label{twist_bc}
\ee
The increase in ground state energy 
$E(\Phi) = \la \psi_\Phi | \hat{H} | \psi_\Phi \ra$
for a small applied twist
\be
E(\Phi)-E(0) = \frac{1}{2} \frac{D_s}{\pi} L^{d} 
\left({\Phi \over L}\right)^2 + \ldots
\label{twist_energy}
\ee
defines the superfluid stiffness $D_s$.
This is related to the superfluid density $n_s$ and the 
helicity modulus $\Upsilon$ \cite{jasnow} via
\be
\frac{D_s}{\pi}=L^{2-d} \lim_{\Phi \rightarrow 0}
\frac{\partial^2 E(\Phi)}{\partial \Phi^2}=2 t n_s = \Upsilon.
\label{defs}
\ee
Note that for a disordered system, the twist definition of $D_s$ 
is valid for any given disorder realization, and thus disorder averaging
can be done at the end of the calculation.

The generalization of the above definitions of $D_s$ to finite temperatures is
straightforward. In the Kubo formula (\ref{jj}) and (\ref{kubo}), the ground
state expectation value has to be replaced by a thermal average, and 
correspondingly in the twisted b.c. formulation (\ref{defs})
the energy has to be replaced by the free energy.

\bigskip

{\noindent {\bf III Bounds from Linear Response Theory}}

\medskip

From (\ref{jj}) we see that $\chi_1({\bf q},\omega=0)$ is non-negative.
It then follows from (\ref{kubo}) that
\be
D_s/\pi \leq \la K_1 \ra.
\label{ke-bd}
\ee
At finite temperatures the right hand side should be viewed as
thermal expectation value.

This bound can be further simplified for the
case of uniform hopping $t_\alpha(\vr) \equiv t$.
Using the Cauchy-Schwarz inequality and the fact that
the geometric mean is less than the arithmetic mean we find
\be
\la K_1 (\vr)\ra \equiv t \la (\crx+h.c.) \ra 
\leq 2t \sqrt{n(\vr) n(\vr + \hat{1})} 
\leq t (n(\vr) + n(\vr+\hat{1})),
\ee
where $n(\vr) \equiv \la c^{\dg}(\vr)c(\vr) \ra$. Further
$\la K_1 \ra \leq t L^{-d} \sum_{\vr} (n(\vr)+n(\vr+\hat{1})) = 2 t \den$
and using (\ref{defs}) we get
\be
n_s \le \la K_1 \ra/2t \le \den.
\label{den-bd}
\ee
Thus, the superfluid density for an interacting lattice or disordered system
can be, and in general is, less than the average particle density. 
The aim of this paper is to understand how this happens and (in the following
Sections) to derive bounds that improve upon the ones above. In the 
present section we focus on the non-disordered problem; we turn to the
the disordered problem in later sections.

Let us begin by emphasizing the importance of broken continuous
translational invariance for $n_s(T=0)$. For a translationally
invariant system, the diamagnetic piece (the analog of $\la K_1 \ra$
in (\ref{kubo})) is given by $\den$. Furthermore the ${\bf q} = 0$ current 
operator is the total momentum which commutes with the Hamiltonian, leading
to a vanishing numerator in (\ref{jj}). Thus for a continuum superfluid 
$\chi_1$  vanishes and $n_s(T=0) = \den$ \cite{pines}
since there are no low lying excitations which can couple to a transverse 
probe. 
(The only low lying excitations present are longitudinal phonons, and even 
these may be gapped for a charged system).
However in the presence of a lattice or impurities, the total (${\bf q} = 0$)
current operator does {\it not} commute with the Hamiltonian in general,
thus $\chi_1$ is non-zero, and (\ref{ke-bd}) is in fact a $strict$ upper bound 
on the stiffness.

It is important to note that approximate treatments 
can miss out on the paramagnetic current correlation piece and
give just the diamagnetic part $even$ on the lattice. For example,
the BCS mean field theory for superconductors reduces the problem
to one of non-interacting particles (Bogoliubov
quasiparticles) with a gapped spectrum and a current operator which
commutes with the reduced BCS Hamiltonian (even on a lattice). 
Thus, estimating the reduction of the superfluid stiffness from its 
diamagnetic value at $T=0$ requires including fluctuations
beyond mean field theory\cite{negU}.

We next discuss interacting (non-disordered) Bose systems on a lattice 
to illustrate the usefulness of the bound (\ref{ke-bd}).
We consider the 2D boson Hubbard model with Hamiltonian
\be
\hat{H}_{\rm bose}=-t \sumrj \left[\crj + h.c. \right]
+ \frac{U}{2} \sum_{\vr} \hat{n}(\vr) \left[\hat{n}(\vr)-1 \right] 
\label{boson}
\ee
where $\hat{n}(\vr)=c^{\dg}(\vr)c(\vr)$.
By comparing the bound with exact results from 
QMC simulations on finite two-dimensional
lattices, we gain insight into how the superfluid stiffness for 
lattice systems depends on the interaction, unlike in the continuum 
case. The calculation of $D_s$ from QMC is described in the
Appendix. Here we summarize relevant numerical details.
The QMC simulations have been done on lattices of size $N=6\times 6$
at commensurate and incommensurate fillings with filling fractions
$\overline{n}=1$ and $\overline{n}=1/2$ respectively.
The inverse temperature $\beta t=4$ is discretised into $N_\tau=32$
steps which implies that $\delta\tau\equiv \beta/N_\tau=1/8$ is the 
elementary time step on which the Trotter approximation is done.
We averaged over $10^8$ MC sweeps through the space-time lattice with 
$10^6$ equilibration sweeps during which we did not collect data. 
The very large number of sweeps
was necessary to get small enough error bars for the superfluid 
stiffness $D_s$ in order to distinguish the
QMC results from the bounds, especially for large interaction
and (or) disorder which is the region of parameter space explored in
both this as well as later sections.
The calculations for each data point took approximately 36 hours on 
a 100 MHz pentium.

In all figures, here and below, we will plot the superfluid stiffness 
normalized such that it is unity for the noninteracting clean (no disorder)
problem at $T=0$, independent of the filling factor. 
We normalize the bounds also by the same factor. Thus,
at $T=0$, what we plot for the stiffness
is really $n_s/\den$ as can be seen from the definitions above since
$n_s=\den$ for the clean noninteracting problem at zero temperature.

First consider the case of incommensurate (non-integral) filling
where the system is expected to be superfluid for all values of the
interaction strength $0\le U \le\infty$.
We see from Fig.~1 that the superfluid stiffness $D_s$, though non-zero
for all $U$, decreases monotonically with the repulsion $U$.
The bound $\pi \la K_1 \ra$ derived in eqn.(\ref{ke-bd})
tracks the behavior of $D_s$ fairly well even at large $U$.
The error
bars on the bounds are much smaller than those on $D_s$ and are hence
not indicated in Fig.~1 or any of the subsequent figures.

Next, consider a commensurate filling of one boson per site (on average).
With increasing $U$ we expect a Mott transition \cite{fisher}
from a superfluid into 
a Mott insulator at a critical value $U=U_c$.
Our aim here is not to locate and characterize this transition very precisely
(which would require a finite size scaling analysis of the QMC data as in
Ref.\cite{KT-euro}). In Fig.~2, we compare the KE bound with
$D_s$ obtained from QMC on the finite lattice.
Even though the bound captures the overall decreasing trend of $D_s$, 
it falls off as $t^2/U$ for large $U$ (as expected from perturbation theory)
rather than vanishing like $D_s$ for roughly $U \geq 20$. 
Being a purely local quantity the KE bound misses the transition.
This brings out the importance of the excitations which
contribute to $\chi_1$ and reduce the stiffness from its diamagnetic value 
eventually driving $D_s$ to zero at the transition. 

It hardly needs to be emphasized that any non-zero upper bound
on $D_s$ such as (\ref{ke-bd}), though valid, will not be useful
for systems which are in fact $non$-superfluid (metallic or insulating).
In such cases the paramagnetic piece $\chi_1$ ignored in the bound 
(\ref{ke-bd}) must exactly cancel $\la K_1 \ra$, leading to $D_s = 0$.
The most trivial example is the non-interacting fermi gas:
the total current operator commutes with the Hamiltonian even on a lattice,
leading to a vanishing numerator in the paramagnetic term in (\ref{jj}), but the
low lying particle-hole excitations contribute in such a way that
there is complete cancellation between the diamagnetic and paramagnetic terms in (\ref{kubo}).
More non-trivial examples are the Mott phase of the boson Hubbard model
discussed above, as well as the large $U$ (fermion) Hubbard model at 
half-filling, which is also a Mott insulator. In both cases,
the total current operator no longer commutes with the Hamiltonian 
and the high energy states (excitations across the charge gap) contribute to
$\chi_1$ leading once again to a complete cancellation in (\ref{kubo}).

\bigskip

{\bf IV. Variational Bounds}

\medskip

We now turn to the task of improving upon the bounds of the
previous Section, especially for disordered systems described by
the Hamiltonian in eqn.(\ref{hamiltonian})
for an interacting Bose or Fermi system in the presence of a
site-dependent disorder potential.
To this end we use a variational method first suggested by Leggett \cite{legg1};
see also \cite{chester}. Although these authors worked in a first
quantized representation, we find it more convenient to work
in the occupation number basis. Let us assume that the
$exact$ (normalized) ground state $\sigr$ of the Hamiltonian 
(\ref{hamiltonian}) with periodic boundary conditions (b.c.) in all
directions is known. We make an ansatz $\sier$ for the ground state 
of (\ref{hamiltonian}) subject to the twisted b.c. (\ref{twist_bc}) 
along the $\hx$ direction, with periodic boundary conditions in the 
other $(d-1)$ directions. Let:
\be
\sier = \exp\left[i \sumall \theta(\vr) \hat{n}(\vr)\right] \sigr
\ee
where $\vr$ runs over all lattice sites and $\theta(\vr)$ are the
variational parameters at our disposal. The twisted b.c. (\ref{twist_bc})
imposes
the constraint
\be
\theta(r_1=L+1,\vr_\perp)-\theta(r_1=1,\vr_\perp)=\Phi. 
\label{constraint}
\ee
where $r_1$ is the coordinate in the $\hx$ direction along which
the twist is applied and
$\vr_\perp \equiv {r_2,\ldots,r_d}$ refers to the set of coordinates
transverse to this direction.
The variational estimate for the difference in energy between the 
twisted and the untwisted boundary condition 
cases is then (for small twist);
\begin{eqnarray}
\Delta E_{\rm trial}[\Phi] & = & \siel H \sier - \sigl H \sigr \nonumber \\
&  = &  \frac{1}{2} \sumrj (\theta(\vr)-\theta(\vr+\hj))^2 
\la K_\alpha(\vr) \ra 
\label{heat}
\end{eqnarray}
where $\la K_\alpha(\vr) \ra = \sigl K_\alpha(\vr) \sigr$.
(In this expression the term linear in $\left[\theta(\vr)-\theta(\vr+\hj)
\right]$
vanishes, since we assume that the ground state $\sigr$
does not break time-reversal symmetry and is thus real
and carries no current on any link.)
Minimizing $\Delta E_{\rm trial}$ with respect to the parameters $\theta(\vr)$ 
(for those $\vr$ $not$ on the boundary
on which the twist is applied) we obtain the following set of
equations:
\be
\sum_{\alpha=1 \ldots d} \{ \la K_{\alpha}(\vr)
\ra \left[\theta(\vr)-\theta(\vr + \hj)\right] + 
\la K_{-\alpha}(\vr) \ra\left[\theta(\vr)-\theta(\vr-\hj)\right]
\} =0 
\label{kirchoff}
\ee
where we have defined 
$K_{-\alpha}(\vr) = t_{-\alpha}(\vr)\left[\crmj + h.c. \right]$
in parallel with (\ref{ke-op}).

There are in all $L^d-L^{d-1}$ equations
implied by eqn.(\ref{kirchoff}) and $L^{d-1}$ equations implied by
the constraint eqn.(\ref{constraint}) thus giving us a total of 
$L^d$ equations which allows us to solve for the phase 
$\theta^{\ast}(\vr)$ at each point of the lattice.
The solution $\{\theta^{\ast}(\vr)\}$ 
of equations (\ref{kirchoff}) and (\ref{constraint})
defines the optimal choice of the local phases in $\sier$
in response to the applied twist.  This in turn
determines $D^{\ast}_s$ via the relation
\be
\frac{1}{2} \frac{D^{\ast}_s}{\pi} L^{d-2}(\Phi)^2
= \Delta E^{\ast}_{\rm trial}
= \frac{1}{2} \sumrj (\theta^{\ast}(\vr)-\theta^{\ast}(\vr+\hj))^2 
\la K_\alpha(\vr) \ra.
\label{var-bd}
\ee
It follows from the variational principle that 
$\Delta E^{\ast}_{{\rm trial}}[\Phi] \ge \Delta E [\Phi]$ and thus 
$D^{\ast}_s \ge D_s$ from eqn.(\ref{var-bd}) above.

It is not possible to start from the set of equations (\ref{kirchoff})
and (\ref{constraint}), and arrive at an analytic closed form solution 
for the bound $D^{\ast}_s$ in general in $d>1$. 
(Later we will discuss the $d=1$ case and also a simplified form which admits
an analytic solution in arbitrary dimension). 
It is however possible
to calculate $D^{\ast}_s$ numerically solving the linear equations 
(\ref{constraint}) and (\ref{kirchoff}).
For this, we first map the problem of obtaining 
$D^{\ast}_s$ on to a particular $random$ $resistor$ $network$ problem 
which gives us some intuition for the results. 
The lattice points $\vr$ of the above system form the nodes of the network,
$\theta(\vr)$ is the voltage at node $\vr$, and 
$\la K_{\pm \alpha}(\vr) \ra$ is the conductance of the link 
$(\vr, \vr \pm \hj)$.  The equations (\ref{kirchoff})
represent (Kirchoff) current conservation at each node $\vr$,
and the constraint (\ref{constraint}) is a constant voltage 
$\Phi$ maintained across this $resistor$ $network$. 
The quantity to be minimised in eqn.~(\ref{heat}) is the (Joule) 
heat dissipation in the network.
Thus $L^{d-2} D^{\ast}_s/\pi$, defined by (\ref{var-bd}), 
is the $conductance$ 
of an equivalent resistor which dissipates the same amount of heat as the 
network for the same voltage difference across its ends. 

It is instructive to write down the solution to the above equations in 
$d=1$ where a closed form result can be obtained. We find 
\be
\theta^{\ast}(x+1)-\theta^{\ast}(x)
={{\Phi}\over{\la K_1(x) \ra}} {{1}\over{\sum_
{x=1\ldots L} {\la K_1(x) \ra}^{-1}}},
\label{1d-twist}
\ee
from which we get
$D^{\ast}_s=L \pi K_{\rm tot}$ with
\be
\frac{1}{K_{\rm tot}}=\sum_{x=1\ldots L}\frac{1}{\la K_1(x) \ra}.
\label{1d-bd}
\ee
It is easy to see that 
$1/K_{\rm tot}$ is simply the total resistance of the $1-$d network
obtained by adding the resistances of the links
in series.
In order to help visualise this result, we show exact
diagonalization results for a system of bosons on a finite
one dimensional lattice in Fig.~3. The Hamiltonian used is
$H_{bose} + \sum_{\vr} V(\vr)\hnr$  where $H_{bose}$ is defined
in eqn.(\ref{boson}).  The parameter values chosen are
given in the figure.  We plot the way the imposed phase
twist at the boundary is accommodated along the chain, and find
that a large phase twist occurs across bonds where the local
kinetic energy is small as expected from eqn.~(\ref{1d-twist}).

The $d=1$ result suggests a simple generalization to arbitrary $d$.
If we assume that $\theta(\vr)$ depends only on $r_1$, the coordinate
along the applied twist direction, and is independent of the transverse 
coordinates $\vr_\perp \equiv \{ r_2,..,r_d \}$ then 
(following the steps in the derivation
of (\ref{var-bd})) we get an explicit analytic expression for the bound 
$D^{o}_s$ as
\be
\frac{D_s}{\pi} \leq
\frac{D^{o}_s}{\pi}= 
L^{2-d}\frac{1}{\sum_{r_1} [\sum_{\vr_\perp} \la K_1(r_1,\vr_\perp) 
\ra]^{-1}}
\label{leg-bd}
\ee
where $\la K_1(r_1,\vr_\perp) \ra \equiv \la K_1(\vr)\ra$.
We note that $D^{o}_s$ is the lattice analog of the
continuum result obtained by Leggett \cite{legg1} and 
Chester \cite{chester}; in the continuum $K_1(\vr)$ gets replaced
by the density $n(\vr)$. The two bounds $D^{\ast}_s$ and $D^o_s$ 
coincide in one dimension as seen from (\ref{1d-bd}) and (\ref{leg-bd}).
The two bounds also coincide for the non-disordered problem in any $d$;
the (discrete) translational invariance ensures that the bound is just 
the mean kinetic energy on the links in the direction of the applied twist. 
However, in general, for $d>1$, $D_{s}^{*}< D^{o}_s$ and is a 
better bound because it has more variational parameters
\cite{continuum}.

We have thus derived all the inequalities in (\ref{results}),
except for the third: $D^{o}_s \leq \pi\la K_1 \ra$. 
The first two inequalities were derived above and the last one 
in (\ref{results}), valid only for systems with uniform hopping,
was derived in (\ref{den-bd}). To complete our task, we use the inequality
$M^{-1}\sum_i a_i \geq [M^{-1}\sum_i a^{-1}_i]^{-1}$
for positive numbers ${a_i}$, $i=1,\dots,M$, in (\ref{leg-bd}).
This yields
\be
\frac{D^{o}_s}{\pi} \leq 
L^{2-d} \frac{1}{L^2} \sum_{r_1} \frac{1}{(\sum_{\vr_\perp} 
\la K_1(r_1,\vr_\perp)\ra)^{-1}}  = \la K_1 \ra.
\ee

\centerline{\bf Generalization to Finite Temperatures}
%%%
\medskip

The above bounds can be easily generalised to finite temperatures using a
trial (variational) density matrix as sketched below.
This can be easily shown to lead to the same equations as the zero
temperature case with the only difference being that we get thermal
expectation values instead of ground state expectation values for the
local kinetic energies. This allows us to compare the bounds with the 
QMC results for $D_s$
in the next section. 
The local kinetic
energies needed for calculating the bounds are also obtained using
QMC as described in the Appendix.

Let $\hat{\rho_0}= \sum_m e^{-\beta E_m} |m\rangle \langle m|$
be the density matrix for the untwisted case,
where $|m\rangle$ denotes the eigenstate of energy $E_m$ for the system
with untwisted boundary conditions.  Consider a trial density matrix 
for the twisted case of the form
\be
\hat{\rho}(\Phi)= 
\exp[{i \sum_{\vr} \hat{n}(\vr)\theta(\vr)}]~\hat{\rho_0}
~\exp[{- i \sum_{\vr} \hat{n}(\vr) \theta(\vr)}].
\label{trial-rho}
\ee
with boundary conditions as in eqn(\ref{constraint}).
The trial free energy is given by
\be
F(\Phi) \leq F_{\rm trial}(\Phi)= $Tr$ [\rhonew \hat{H}] +
\beta^{-1}~{\rm Tr} \left[\rhonew {\rm ln} \rhonew \right]
\ee
and we need to minimize $F_{\rm trial}(\Phi)$ as before. 
Using the unitary nature of the transformation in eqn.(\ref{trial-rho}) and
the cyclic property of the trace,  
it is straightforward to show that the entropy term 
$\beta^{-1}~$Tr$ [\rhonew $ln$ \rhonew]$ 
remains unchanged even when the twist is put in. The energy term 
Tr $(\rhonew \hat{H})$ leads to an expression identical to
that obtained at $T=0$ with thermal expectation values instead.

\bigskip

\noindent
{\bf V. Comparison with Quantum Monte Carlo Results:}

\medskip

In this Section we illustrate the usefulness of the bounds
by comparing them with the stiffness $D_s$ obtained from QMC 
\cite{KNC,Batrouni}
for interacting Bose systems 
described by the Hamiltonian in eqn.(\ref{hamiltonian}) with
$t_{\alpha}(\vr)=t$ and
\be
\hat{V}={{U}\over{2}} \sum_{\vr} \hnr (\hnr-1) + \sum_{\vr} V(\vr) \hnr
\label{boson-v}
\ee
We work in two dimensions and choose the site dependent (disorder)
potential $V(\vr)$ as given below for various cases.

We first consider a particular case of broken (discrete) 
translational invariance to illustrate the difference between the 
various bounds in (\ref{results}).
We choose a certain set of sites on the lattice
where $V(\vr) = V > 0$ (marked in the inset to Fig.~4)
while fixing the potential at other sites to be zero. 
We then study the effect of varying $V$ on the superfluid 
stiffness $D_s$ and compare it with the various bounds, 
all computed for a twist applied along the $x-$direction. 
The results are shown in Fig.~4: as $V$ is increased, the mean KE bound 
and $D^o_s$ are both considerably larger than the true stiffness 
while the bound $D^{\ast}_s$ tracks the true stiffness extremely well. 
For instance at large $V$, as seen from Fig.4, 
$\la K_1 \ra \sim 0.8$, $D^o_s/2\pi \overline{n}t \sim 0.5$, 
while $D^{\ast}_s/2\pi\overline{n}t$ and $D_s/2\pi
\overline{n}t$ are both $\sim 0.05$ and
thus $D^{\ast}_s$ is about an order of magnitude better than the
other bounds.
We do not know if the excellent agreement between 
$D^{\ast}_s$ and $D_s$ would persist in the
thermodynamic limit, nevertheless it is clear
that $D^{\ast}_s$ does considerably better than the other bounds.

We next consider the disordered Boson Hubbard model, which is defined 
by (\ref{boson-v}), with $V(\vr)$ at each $\vr$ an independent random 
variable uniformly distributed in the interval 
$[-V_{\rm dis},V_{\rm dis}]$.
We study the case of incommensurate filling, i.e., non-integral
average site occupancy. As mentioned earlier, the bounds are valid
for any given disorder realization.  We therefore choose to consider a
particular realization of the random potential in Figs.~5 and 6
in order to illustrate the bounds.

In Fig.~5 we keep the disorder strength 
$V_{\rm dis}$ fixed, and look at the interaction $U$-dependence
of the superfluid stiffness and the bounds. For small $U$, 
we see that $D_s$ is small, since at $U=0$ we expect that 
the system is in a localized phase, with all bosons in the 
lowest lying, localized single-particle eigenstate. Note
that due to the finite size of the calculation, $D_s$ does not
vanish even at $U=0$. Nevertheless, the bounds $D^{\ast}_s$ and
$D^o_s$ do capture the large suppression of $D_s$ at small $U$,
while the mean kinetic energy is clearly unable to do so. 
For intermediate $U$ we find nonmonotonic $U$-dependence in $D_s$
\cite{KNC}. 
From a mathematical point of view, there is no guarantee that because
$D_s$ is a nonmonotonic function of some parameter, say $U$, an upper
bound on $D_s$ should show similar $U$-dependence. 
It is therefore interesting that this feature $is$ captured by 
$all$ the bounds.  At larger values of $U$, 
$D^{\ast}_s$ and $D^o_s$ are closer to the KE in value, 
than to $D_s$, but all three track the overall trend of $D_s$ fairly well. 

In Fig.~6 we keep the interaction $U$ fixed and study $D_s$ and the bounds
as a function of disorder strength $V_{\rm dis}$. We find that $D_s$
decreases monotonically with disorder. This trend is captured by all
three bounds, however, we clearly see that
$D^{\ast}_s$ does increasingly better than the other bounds in the 
large disorder regime.

\bigskip

\noindent
{\bf VI. Remarks:}

\medskip

In this Section we collect together some remarks
which are meant to clarify different points which we
feel are important. We also consider a model where the bound
$D^{\ast}_s$ captures a transition unlike in the examples presented earlier.

\medskip

\noindent
{\bf Charge Stiffness versus Superfluid Stiffness}

\smallskip

A bound similar to the mean kinetic energy bound (\ref{ke-bd})
has been discussed in Ref.~\cite{millis} in connection with the 
charge stiffness, which is the coefficient of $\delta(\omega)$
in the optical conductivity of a perfect conductor.
We emphasize that although the linear response formalism leads to 
essentially identical bounds for both the charge stiffness and the superfluid 
stiffness $D_s$, the variational wavefunction construction of Section IV
and the associated bounds are valid only for $D_s$.
Recall that the charge stiffness is defined in terms of the energy of 
the adiabatically generated ground state as the twist is introduced 
\cite{swz}.
The variational ansatz that is considered here is by construction
a trial ground state of the system at finite twist and is {\em not}
an adiabatic continuation of the untwisted ground state. 
We therefore emphasize that unless there is a gap which makes the
charge and superfluid stiffness equal \cite{swz}, the above variational
construction {\it does not} give us a bound on the charge stiffness.

\medskip

\noindent
{\bf Kinetic Energy : Exact versus Approximate Estimates}

\smallskip

We should note that our bounds are expressed in terms of the local
kinetic energies evaluated in the {\it exact} ground state
which is in general not known.  Nevertheless, one can derive useful 
qualitative information simply from the form of the bounds.
In order to get quantitative estimates of the bounds and of $D_s$
one needs to resort to numerical methods like QMC. However, the bounds
which involve local correlations, are much less affected by finite
size errors. 

If a trial ground state wave function is used, instead of the exact one, 
to evaluate the local kinetic energies, then the expressions
derived above do $not$ lead to rigorous upper bounds. 
Such approximations could, nevertheless, lead to useful estimates.

\medskip

\noindent
{\bf Quantum Percolation Model}

\smallskip

We now describe an example which highlights a qualitative difference 
between the bounds $D^{\ast}_s$ and $D^o_s$ and where $D^{\ast}_s$
in fact captures a phase transition unlike earlier examples of both clean
and disordered systems. Consider a ``quantum bond 
percolation model'' of bosons on a lattice with hopping disorder such
that the hopping matrix element between two neighbouring sites is either
$t$, with probability $(1-p)$, or zero with probability $p$. 
We assume that the twist is applied along a certain direction and 
that we have fixed boundary conditions in the perpendicular directions.

As the probability $p$ of having a ``broken link'' increases, 
the superfluid stiffness reduces until it vanishes at some 
$p=p_{\rm loc}$ via a (quantum) $localization$ $transition$. 
We do not expect either $D^{\ast}_s$ or $D^o_s$ to show a signature
of this transition. The bound $D^{\ast}_s$ would
go to zero when there is $any$ set of broken links that can break up
the lattice into two disconnected parts across the twist direction.
Thus we expect $D^{\ast}_s$ to vanish at $p=p_c$ corresponding to the 
{\it classical bond percolation transition} and $p_c > p_{loc}$.
In order for $D^o_s$ to vanish, the lattice needs to be disconnected
along a (hyper)plane {\it perpendicular to the twist direction}. 
Since the hopping matrix elements are chosen randomly, 
realizations with such a set of broken links will be of measure zero.
Thus $D^o_s$ will not capture this percolation transition.

\medskip

\noindent
{\bf Bounds on $T_c$ in 2D:}

\smallskip

In two dimensions an upper bound on the superfluid density 
allows us to put an upper bound on the superfluid to normal transition
temperature $T_c$ in a clean system. For this we use the relation between
the jump in superfluid density at the Kosterlitz-Thouless (KT) transition and
the KT transition temperature \cite{nelson}
$D_s(T^{-}_c)=2 T_c$ (with $k_B=1$). Thus,
$D_s(T^{-}_c) \leq D^{\ast}_s(T_c) \leq D^{\ast}_s(T=0)$ 
which gives us the following bound on $T_c$:
\be
T_c \leq D^{\ast}_s(T_c)/2 \leq D^{\ast}_s(T=0)/2.
\ee
Note, however, that in general a lower superfluid density at 
zero temperature does not necessarily imply a lower $T_c$. 
A counterexample is the case of s-wave BCS superconductors with 
weak disorder: Anderson's theorem tells us that in this case $T_c$ 
and other thermodynamic properties are 
unaffected but one would expect $D_s$ to be suppressed by the disorder.

\bigskip

\noindent
{\bf VII. Conclusions}
\medskip

In this paper we have obtained several upper bounds on the superfluid 
stiffness at zero temperature as well as finite temperatures. These 
have been obtained using the Kubo linear response formalism as well as
variational estimates. These bounds give 
qualitative insight into the dependence of $D_s$ on various parameters. 

For clean systems, the average kinetic energy $\la K_1 \ra$ gives a good 
upper bound on $D_s/\pi$ and provides insight into the difference between 
lattice and translationally invariant (continuum) models.  The bound 
captures the qualitative trend of the
superfluid stiffness $D_s$ to decrease with increasing repulsive interaction
$U$. For incommensurate filling, the bound compares well with $D_s$ at all
values of the interaction. At commensurate filling, the bound is good at
small values of $U$ but fails to capture the superfluid-Mott insulator 
transition at large $U$.

For disordered 
systems we obtain two improved variational bounds $D^{\ast}_s$ and
$D^o_s$ ($D^{\ast}_s\le D^o_s \le \pi\la K_1\ra$)
by allowing larger phase twists on links with lower
local kinetic energy. We quantitatively compare the bounds with QMC results 
and find that they capture non-trivial parameter dependence of $D_s$, 
including its nonmonotonic variation in some cases.

\bigskip

\bigskip

\vfill \eject

\noindent
{\bf Appendix }

\medskip

{\bf Path Integral Quantum Monte Carlo}

\smallskip

In this Appendix we briefly describe the path integral QMC method used to 
calculate the local kinetic energies and the superfluid stiffness $D_s$ 
presented in the paper. For more details see Refs.~\cite{cep-RMP,NT-book}.
The notation in this appendix differs from that in the rest of the paper
since here we work in first quantized representation. 
The position vector of the $n^{th}$ boson, $n=1 \ldots N_b$,
at imaginary time $\tau$, where $0 \le \tau \le \beta$, 
is denoted by $\vr_{n}(\tau)$. 
The diagonal density matrix for bosons in the canonical ensemble 
is given by
\be
\rho(R,R;\beta)={{1}\over{N_b !}}
\sum_{\cP}\la R|\exp(-\beta H) |\cP [R]\ra 
\ee
which 
describes the amplitude for the $N_b$-particle system starting in 
a configuration $R=\lbrace \vr_1, \vr_2, \ldots, \vr_{N_b}\rbrace$ at 
time $\tau=0$ to be in configuration $\cP[R]$, where $\cP$ is a 
permutation of $R$, at time $\tau=\beta=1/T$. 
The partition function is given by $Z=\sum_{R} \rho(R,R;\beta)$.
Each particle describes a path or a world-line in the $\vr$-$\tau$ 
space, and the evaluation of the partition function amounts to summing 
over all possible paths which is most efficiently 
done by QMC techniques. 

The first step in evaluating the density matrix is to break the total
time $\beta$ into small steps $\dt=\beta/N_\tau$. Upon inserting complete 
sets of states we get
\begin{equation}
\rho(R,R;\beta) = {1\over {N_b !}} \sum_{\cP} \sum_{R_1}\cdots
		\sum_{R_{(N_\tau -1)}}
	\rho(R,R_1;\dt) \rho(R_1,R_2;\dt) \cdots 
	\rho(R_{(N_\tau-1)},\cP[R];\dt)
\label{dm}
\end{equation}

A particular configuration of the particles on all the time slices,
is denoted by
\be
\cS = [\cP; R,R_1,R_2,\ldots R_{(N_\tau -1)}, R_{N_\tau}]
\ee
and looks like a collection of strings. Our aim is to sample the 
probability function
\be
{\rm Prob}[\cS] = \rho(R,R_1;\dt) \rho(R_1,R_2;\dt) \cdots 
	\rho(R_{(N_\tau-1)},\cP[R];\dt)
\label{prob}
\ee
We use the Metropolis importance sampling method to generate
a sequence of M string configurations $\cS_1,\cS_2,\ldots \cS_M$
such that in the limit $M\rightarrow \infty$ the probability of finding a
particular configuration $\cS_\mu$ in this sequence is 
proportional to ${\rm Prob}[\cS_\mu]$. 
The actual implementation of the Monte Carlo method, including importance 
sampling, and the bisection method for efficient exploration of phase 
space, is described in Ref\cite{NT-book}. 

\bigskip

{\bf Calculation of the Local Kinetic Energy}

\medskip

For simplicity, first consider 
an operator $O=\sum_i f(\hat{n}_i)$ which is diagonal in the position 
basis, like the interaction energy in the second term of eqn.(\ref{boson})
(similar considerations apply to the disorder energy in eqn.(\ref{boson-v})).
Then 
\be
\la O \ra \equiv {{{\rm Tr} O e^{-\beta H}}\over{{\rm Tr} e^{-\beta H}}}
={1\over M} \sum_{\mu=1}^M {1\over {N_\tau}}\sum_{\tau=1}^{N_\tau}
\sum_i f(n^{\mu}_{i}(\tau))
\label{potl_en}
\ee
where $n^{\mu}_i(\tau)$ is the number of bosons at site $i$ on time slice 
$\tau$ in the string configuration $\cS_\mu$.

Next, consider the local kinetic energy operator given in eqn.(\ref{ke-op}).
Recall that in the Monte Carlo procedure,
the density matrix in an elemental time $\dt$ is approximated by
\begin{equation}
\rho(R_\tau,R_{\tau+1}) = \prod_{n=1}^{N_b}
	\rho_1(\vr^\mu_n(\tau),\vr^\mu_n(\tau+1))
	\times \exp\left[-(\dt/2)\sum_j [{\cal U}^\mu_i(\tau) + {\cal U}^\mu_i 
						(\tau+1)]\right ]
\label{dm_tau}
\end{equation}
where $\rho_1$ is the single particle density matrix corresponding to the
kinetic energy and disorder potential operators and 
${\cal U}^\mu(n_i(\tau)) = (U/2) n^\mu_i(\tau) (n^\mu_i(\tau)-1)$ arises from 
the interaction between the bosons.

To evaluate the kinetic energy, pick a string configuration 
$\cS_\mu$ with the $n^{th}$ boson at $\vr^\mu_n(\tau)$ on time slice $\tau$
and imagine moving that boson to one of its four neighboring 
sites $\vr^{\prime\mu}_n (\tau)$
on the same time slice generating a new configuration. 
For clarity, we suppress the $\mu$ index labelling a
configuration in the following equations. 
The change in potential energy is given by ${\cal U}_{new} - {\cal U}_{old}$. 
\begin {equation}
U_{old} = {\cal U}\left[n(\vr_n(\tau))\right] 
+ {\cal U}\left[n(\vr^\prime_n (\tau))\right]
\label{uold}
\end{equation}
and
\begin {equation}
U_{new} = {\cal U}\left[n(\vr_n(\tau))-1\right] + 
	{\cal U}\left[n(\vr^{\prime}_n(\tau))+1\right] 
\label{unew}
\end{equation}
There will also be a change in the single particle density matrix.

Thus the local kinetic energy for a given string configuration $\cS_\mu$ is
\begin{equation}
K(\vr_n(\tau), \vr^{\prime}_n(\tau)) = 
{{\rho_1(\vr^{\prime}_n (\tau),\vr_n(\tau+1)) } \over
{\rho_1(\vr_n (\tau),\vr_n(\tau+1)) }}
 \exp[-U_{new} + U_{old}]
\label{ke2}
\end{equation}
which is then averaged over all the configurations selected by the 
Monte Carlo procedure.

\bigskip

{\bf Calculation of Superfluid Stiffness $D_s$}

\medskip

The current along the $\alpha^{th}$ direction is given by
\begin{equation}
j_\alpha(\tau) = \sum_{n=1}^{N_b}
[r_{n\alpha}(\tau+\delta\tau)-r_{n\alpha}(\tau)]/\delta\tau
\label{current}
\end{equation}
where  $r_{n\alpha}(\tau)$ is
the $\alpha^{th}$ component of $\vr_n(\tau)$ ($\alpha=1 \ldots d$).
The winding number $W_\alpha$ which counts the number of times the path
of a particle winds around a box with periodic boundary conditions is
given by \cite{cep-pol}
\be
W_\alpha = {1\over {N_\tau}} 
\sum_{\tau=1}^{N_\tau} j_\alpha (\tau)
\ee
and the superfluid stiffness is given by
\be
{{D_s}\over{\pi}} = L^{2-d} \sum_{\alpha} \la  W^{2}_{\alpha} \ra
\label{qmc-ds}
\ee
In the case of Fig.4, we calculate the winding number, the stiffness
and the bounds all along the $x-$ direction only as indicated in the 
figure inset. For all other figures, we calculate the total $D_s$ given
in eqn.(\ref{qmc-ds}).

%\end{multicols}

\bigskip

\vfill \eject

\vfill \eject

\begin{figure}
\vspace{0pt}
{\epsfxsize=3.0truein\epsfbox{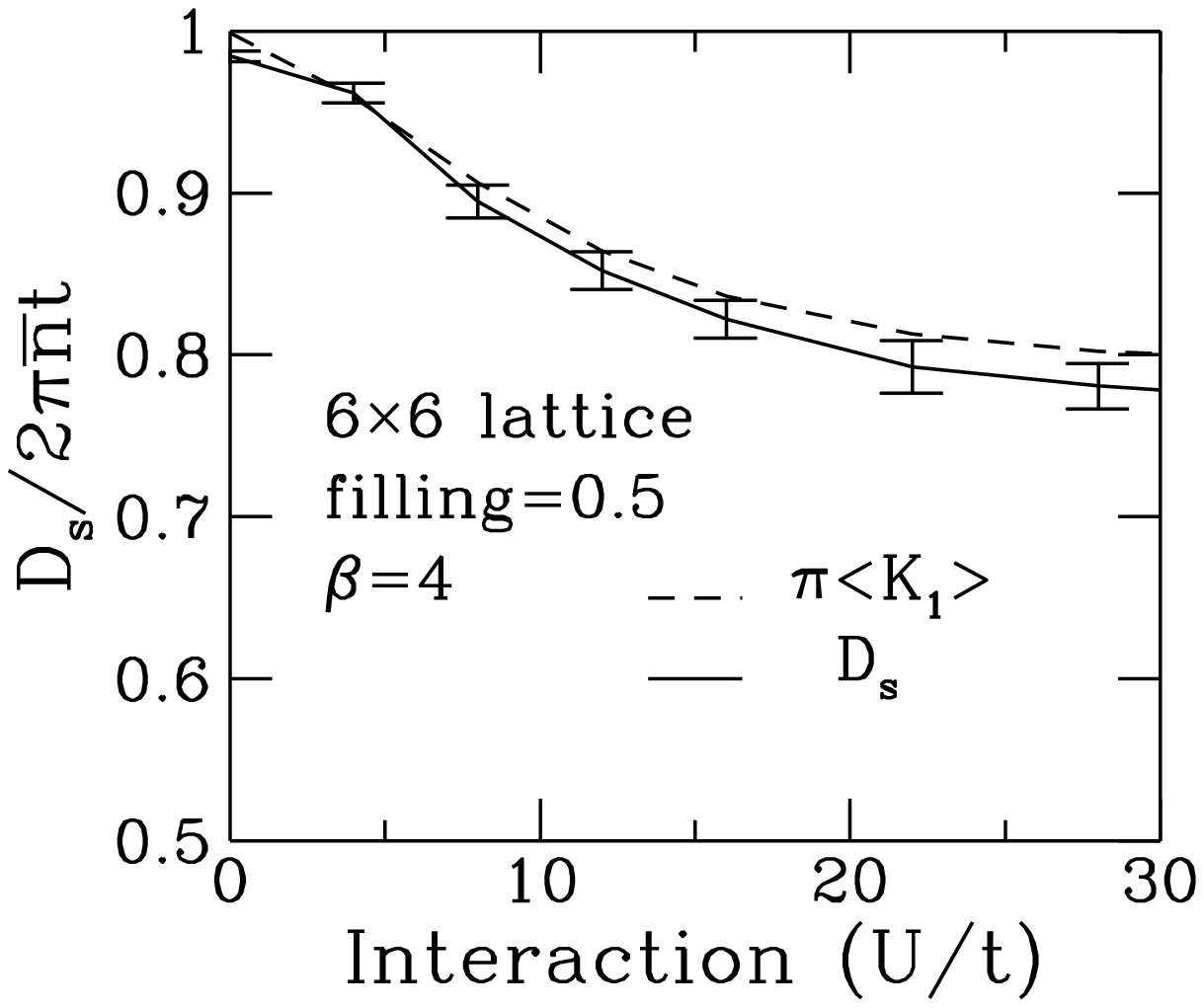}}
\vskip 0.5truecm
\caption{The superfluid stiffness $D_s$ and the KE bound 
$\pi \la K_1 \ra$ obtained from path integral QMC for the Boson Hubbard 
model in eqn.~(13) at an
incommensurate filling $\bar{n}=0.5$ for parameter values shown in the 
figure.
The QMC data here (and in other figures) was averaged over $10^8$ runs
to sufficiently reduce error bars in order to allow comparison with the
bounds. The normalization here and in the remaining figures is chosen to
make the normalized stiffness unity for a clean non-interacting system
at $T=0$ independent of the filling factor.
}
\label{1}
\end{figure}

\begin{figure}
\vspace{0pt}
{\epsfxsize=3.0truein\epsfbox{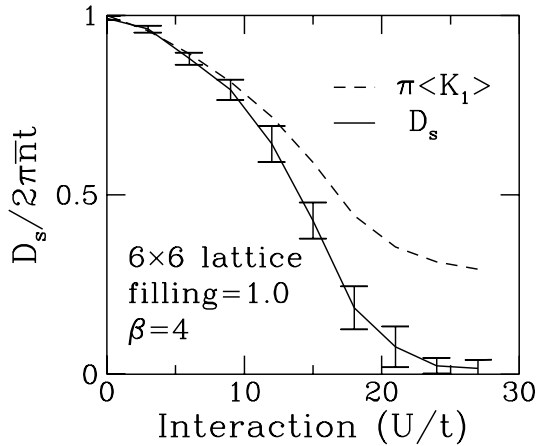}}
\vskip 0.5truecm
\caption{ Superfluid stiffness $D_s$ and KE bound $\pi\la K_1\ra$ for
the Boson Hubbard model 
(13) 
plotted for commensurate filling
$\bar{n}$ and parameter values indicated in the figure.
In the thermodynamic limit, $D_s$ would vanish for $U>U_c$ (where
$U_c/t \sim 20 $) signalling the superfluid to Mott insulator transition
while on a finite system it
remains non-zero even beyond that value and is zero (within error bars)
only at large $U$.
The bound however is clearly non-zero even at large $U$ and can be shown
from perturbation theory
to go as $t^2/U$ for $t>>U$. The bound {\em does not}
capture the Mott transition.
}
\label{2}
\end{figure}

\vfill\eject

\begin{figure}
\vspace{0pt}
{\epsfxsize=3.5truein\epsfbox{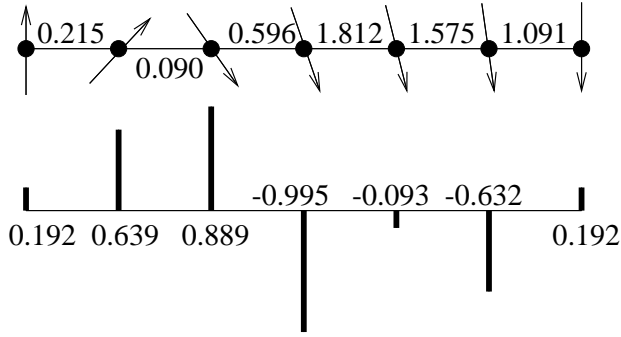}}
\vskip 0.5truecm
\caption{The phase $\theta(\vr)$ calculated at each site 
(shown as arrows) indicating how the
applied phase twist $\Phi=\pi$ is accommodated along the lattice in a 
one-dimensional case for a particular realization of disorder for the
dirty boson problem of eqn.~(25). The absolute 
value of the kinetic energy is 
indicated along the links and the disorder potential
is shown in the lower 
part of the figure (both in units of $t$.) 
The ``bad bonds'' (the first two links with
smaller kinetic energy) allow for a greater phase twist across them as shown.
The data was obtained by exact diagonalization for a system of 4 bosons
on a 6-site chain.}
\label{3}
\end{figure}

\begin{figure}
\vspace{0pt}
  {\epsfxsize=3.0truein\epsfbox{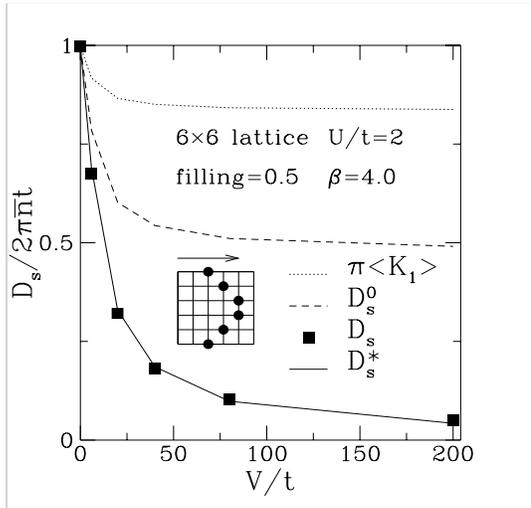}}
  \vskip 0.5truecm
  \caption{
  Superfluid stiffness plotted for a $6\times 6$ system at
  incommensurate filling in which the sites indicated in the inset
  are kept at a potential $V$ with the rest kept at a potential of zero.
  The phase twist is applied along the direction shown by the arrow in the
  inset
  and the bounds and the stiffness $D_s$ are obtained only along that
  direction. The error bars on $D_s$ are of the size of the data point
  and hence not indicated. The difference between the various bounds is 
  evident. Notice that the bound $D^{\ast}_s$
  is better than the other bounds by about an order of magnitude.
  }
\label{4}
\end{figure}

\vfill\eject

\begin{figure}
\vspace{0pt}
  {\epsfxsize=3.0truein\epsfbox{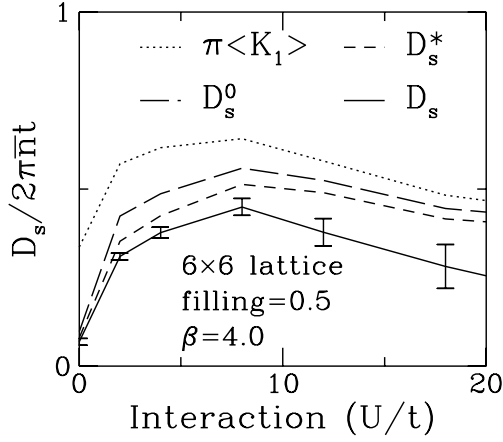}}
  \vskip 0.5truecm
  \caption{
  Superfluid stiffness plotted as a function of interaction $U$ for the
  dirty boson
  problem in eqn.~(25) at incommensurate filling for $V_{dis}/t=10$
  and other parameter values indicated in the figure.
  The bounds capture the non-monotonic behavior of $D_s$
  obtained from QMC as a function of $U$.
  The bound $D^{\ast}_s$ is clearly better than the KE bound when
  compared with $D_s$.
  At large $U$, the kinetic energies on various links are more
  uniform and the bounds are closer to each other.
  }
\label{5}
\end{figure}

\begin{figure}
\vspace{0pt}
  {\epsfxsize=3.0truein\epsfbox{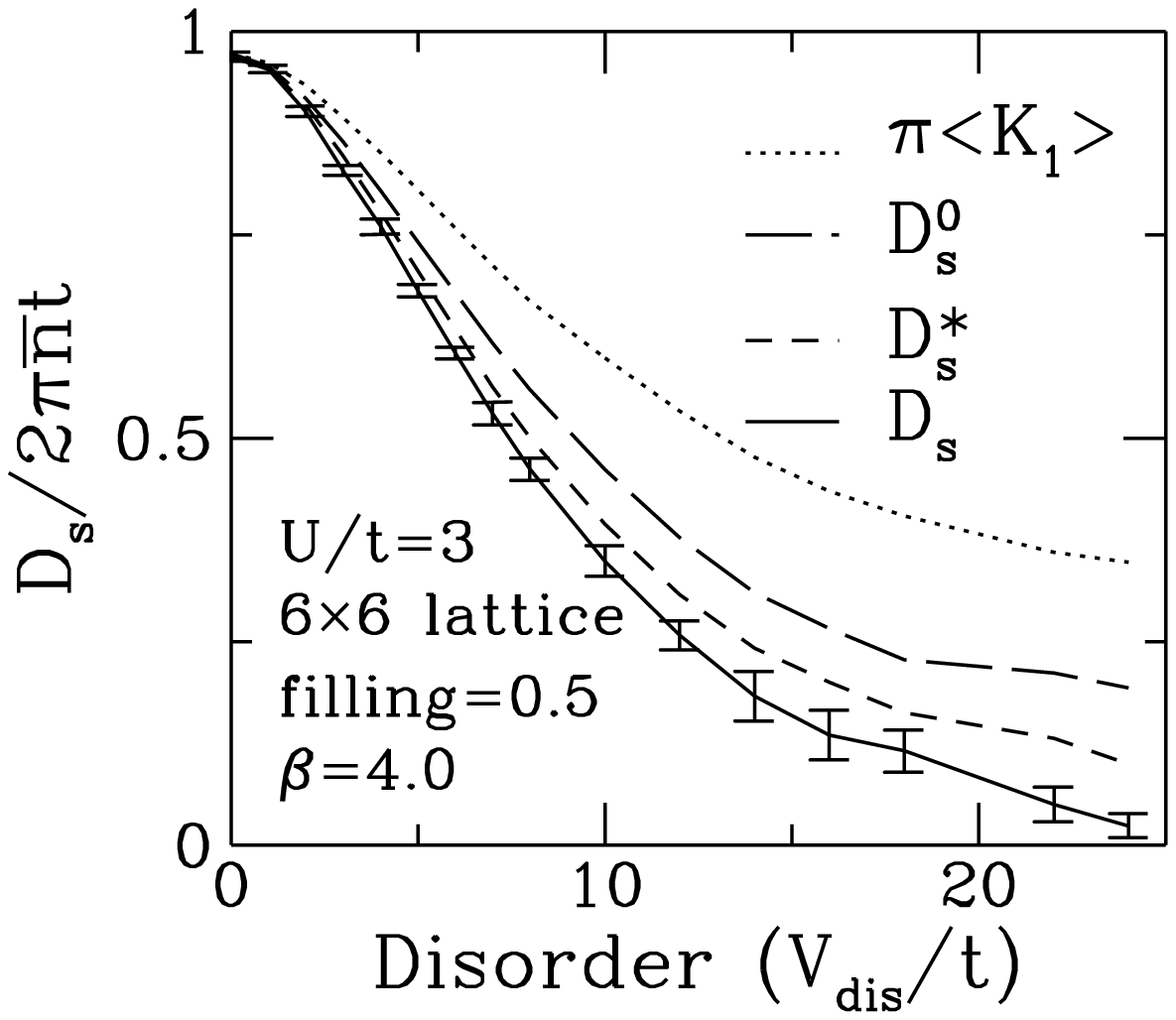}}
  \vskip 0.5truecm
  \caption{
  Stiffness plotted as a function of disorder, $V_{dis}$, for the
  dirty boson problem in eqn.~(25) at incommensurate filling
  for parameter values indicated in the figure.
  The bound $D^{\ast}_s$ gets progressively better than the other two bounds in
  the disorder dominated regime. This is precisely the regime where the local
  kinetic
  energies are very inhomogeneous.
  }
\label{6}
\end{figure}


\begin{thebibliography}{999}
\bibitem{continuum}
The inequality $D^{\ast}_s < D^o_s$ continues to hold in the
continuum case as well, with obvious changes like $K_1(\vr)$ being
replaced by the density $n(\vr)$. The bound $D^{\ast}_s$ should be useful for 
continuum systems with disorder and also for systems, such as the
one considered in Ref.~\cite{legg1}, 
in which translational invariance is spontaneously broken.
\end{thebibliography}
\end{document}